\documentstyle{article}
\def\1ad{\mbox{\normalsize $^1$}}
\def\2ad{\mbox{\normalsize $^2$}}
\def\3ad{\mbox{\normalsize $^3$}}
\def\4ad{\mbox{\normalsize $^4$}}
\def\5ad{\mbox{\normalsize $^5$}}
\def\6ad{\mbox{\normalsize $^6$}}
\def\7ad{\mbox{\normalsize $^7$}}
\def\8ad{\mbox{\normalsize $^8$}}
\def\makefront{
\begin{flushright}
SPIN-1999/02\\ 
hep-th/9903003
\end{flushright}
\vskip  0.5truecm
\begin{center}
\def\newtitleline{\\ \vskip 5pt}
{\Large\bf\titleline}\\
\vskip 1truecm
{\large\bf\authors}\\
\vskip 5truemm
\addresses
\end{center}
\vskip 1truecm
{\bf Abstract:}
\abstracttext
\vskip 1truecm}
\newcommand{\be}{\begin{equation}}
\newcommand{\ee}{\end{equation}}
\newcommand{\bea}{\begin{eqnarray}}
\newcommand{\eea}{\end{eqnarray}}
\newcommand{\nn}{\nonumber \\}

\newcommand{\ba}{\begin{array}}
\newcommand{\ea}{\end{array}}

\renewcommand{\a}{\alpha}

\def\bbox{{\,\lower0.9pt\vbox{\hrule \hbox{\vrule height 0.2 cm
\hskip 0.2 cm \vrule height 0.2 cm}\hrule}\,}}
\newcommand{\dsl}{\pa \kern-0.5em /}

\newcommand{\pa}{\partial}

\font\mybb=msbm10 at 12pt
\def\bb#1{\hbox{\mybb#1}}

\def\bE {\bb{E}}

\def\a{\alpha}\def\g{\gamma}

\def\a{\alpha}

\def\f{\phi}               
\def\g{\gamma}

\def\cn{{\cal N}}

\def\car{{\cal R}}

\begin {document}
\def\titleline{
Field theory limit of branes and gauged supergravities
}
\def\authors{
Kostas Skenderis
}
\def\addresses{
Spinoza Institute, University of Utrecht, \\
Leuvenlaan 4, 3584 CE Utrecht, The Netherlands\\
{\tt k.skenderis@phys.uu.nl}
}
\def\abstracttext{
We discuss the field theory limit of D$p$-branes. 
In this limit, the black D$p$-brane solution approaches a solution 
which is conformal to $adS_{p+2} \times S^{8-p}$. 
We argue that the frame in which the conformal 
factor is equal to one, the dual frame, is a `holographic' frame.
The radial coordinate of $adS_{p+2}$ provides a UV/IR 
connection  as in the case of the $D3$ brane.
The gravitational description involves gauged supergravities,
typically with non-compact gauged groups. The near-horizon D$p$-brane solution 
becomes a domain-wall solution of the latter.
}
\large
\makefront
Holography\cite{holography} states that a gravitational system 
in $d+1$-dimensions should have a description in terms of a 
$d$-dimensional (boundary) field theory.
In addition, the boundary theory should not contain more than 
one degree of freedom per Planck area. 
The adS/CFT duality\cite{malda,GKP,Wit1} provides
an example of such holographic connection\cite{Wit1,SW}. For instance,
anti-de Sitter supergravity in five dimensions has a description
in terms of (strongly coupled) $\cn=4$ $SU(N)$ SYM theory in four dimensions.
This duality was inferred by looking at two descriptions 
of the D3 brane: one as a black D3 brane and another 
as a hypersurface where strings can end. Taking the field theory
limit, i.e. the limit in which the bulk gravity decouples,
one finds that the worldvolume theory is equivalent to 
strings propagating in the near-horizon limit of the black D3 brane
which is $adS_5 \times S^5$. When curvatures are small the YM
coupling constant is strong and we obtain that  
anti-de Sitter supergravity is equivalent to strongly coupled 
SYM theory.

It is natural to consider the same limit for the other branes as 
well. The difference between the D3 brane and the other branes
is that the worldvolume theory of the latter is not conformal.
Therefore, the dual supergravity cannot be anti-de Sitter supergravity.
By holography we expect that when the YM coupling constant becomes
large a gravitational description in one dimension higher 
takes over. Indeed we will see that this is the case\cite{BST}: 
The gravity description is in terms of certain gauged supergravities in 
$p+2$ dimensions, typically with non-compact gauge groups, which 
possess supersymmetric domain-wall vacua. These domain-wall 
vacua are spacetimes conformal to $adS_{p+2}$ 
(for $p=5$ we get $\bE^{(1,6)}$ instead).

Let us consider the field theory limit \cite{malda,maldatwo} of D-branes.
We want to consider a limit in which the bulk gravity decouples
and we left with a decoupled worldvolume theory. To decouple
closed string loop effects we send $g_s \to 0$. To suppress
higher dimension operators we go to low energies, $\a' \to 0$. This implies
that the gravitational coupling constant, i.e. Newton's constant 
$G_N \sim \a'^4 g_s^2$, is also sent to zero, and therefore 
gravity decouples (there are subtleties with the decoupling 
limit in the case of $D6$ branes that we do not discuss here). 
We want to take the limit in such way that
there is still a non-trivial worldvolume theory.
Since the worldvolume dynamics are due to 
open strings, we keep the masses of stretched strings, $U=r/\a'$, fixed as 
we go to low energies. In addition, we keep $g_{YM}^2 N$ fixed, where 
$g_{YM}^2 \sim g_s \a'^{(p-3)/2}$ is the YM coupling constant. 
We choose to include a factor of $N$ because the open string coupling 
includes a factor of $N$ (since strings can end in any of the $N$ 
D$p$-branes). In addition,
it is $g_{YM}^2 N$ that appears naturally in the gravitational
description. To summarize, we consider the limit
\be \label{limit}
g_s \to 0, \qquad \a' \to 0, \qquad U={r \over \a'}=\mbox{fixed}, \qquad
g_{YM}^2 N=\mbox{fixed}
\ee
This means that $g_s N \to 0$ for $p<3$, $g_s N$=fixed for $p{=}3$
and $g_s N \to \infty$ for $p>3$. 

The solution that describes a black D$p$-brane, in the string frame, 
is given by
\bea\label{pbrane}
&&ds^2_{st} = H^{-1/2} ds^2(\bE^{(p,1)}) + H^{1/2}ds^2(\bE^{(9-p)})]\nn
&&e^\phi = g_sH^{(3-p)/4}, \nn
&&F_{8-p} = g_s^{-1} \star dH,
\eea
where $*$ is the Hodge dual of $\bE^{(9-p)}$ and 
$H = 1 + g_sN(\sqrt{\a'}/r)^{(7-p)}$. 

In the limit (\ref{limit}) the harmonic function tends to 
\be \label{Hlimit}
H= 1 + {g_{YM}^2N\over (\alpha')^2 U^{(7-p)}} \rightarrow
g_{YM}^2N(\alpha')^{-2} U^{(p-7)}.
\ee
By holography we expect that in this limit 
the geometry would factorize into a $(8-p)$ compact 
manifold times a solution of $p+2$-dimensional supergravity.
Inserting (\ref{Hlimit}) into (\ref{pbrane}), however, 
we find that this is not the case. Moreover, the resulting solution 
is singular. 

A closer look at the solution reveals that it is conformal to 
$adS_{p+2} \times S^{8-p}$\cite{BPS1,BST} (except for $p=5$ in which case 
one gets $\bE^{(1,6)} \times S^3$). Therefore,
the problems mentioned above are due to improper choice of 
frame. The frame in which the solution becomes exactly 
$adS_{p+2} \times S^{8-p}$ is the so-called dual-frame.
This is defined\cite{DGT} 
as the frame in which the dual field strength 
$F_{8-p}$ couples to the dilaton the same way as the metric.
This means that the dual metric is related to the string metric as
\be \label{Dmetric}
g_{dual} = (e^{\phi}N)^{{2 \over p-7}} g_{st}.
\ee
In this frame the relevant part of the action is  
\be \label{action}
S = {N^2\over \a'{}^4} 
\int d^{10}x\, \sqrt{-g} (N e^{\phi})^\gamma \big[ R + {4(p-1)(p-4)\over
 (7-p)^2} (\partial \phi)^2 - {1\over 2 (8-p)!}\, 
{1 \over N^2}\, |F_{8-p}|^2 \big]
\ee
where $\gamma = 2(p{-}3)/(7{-}p)$.
Let us define the coordinate (for $p \neq 5$)
\be \label{newvar}
u^2 = \car^2 (g_{YM}^2 N)^{-1} U^{5-p} \qquad [\car=2/(5-p)]\, .
\ee
Then the D$p$-brane solution takes the form 
\bea\label{compac}
&&ds^2_{dual} = \a'\left[\frac{u^2}{\car^2} ds^2(\bE^{(p,1)}) 
+ \car^2 \frac{du^2}{u^2}
+ d\Omega^2_{(8-p)}\right]\nn
&&e^{\phi} = {1 \over N} 
(g^2_{YM} N)^{{(7-p) \over 2(5-p)}}
({u \over \car})^{{(p-7)(p-3) \over 2(p-5)}}, \nn
&&F_{8-p} = (7-p) N (\a')^{{(7-p)\over 2}} vol(S^{8-p}).
\eea
The metric is that of $adS_{p+2} \times S^{8-p}$. 

The dual frame may be considered as a ``holographic frame'' since it is in this
frame that the black D$p$-brane solution immediately leads, 
in the decoupling limit, to a $p+2$ gravitational description. 
Furthermore, as we now show the area of each
horosphere (i.e. a hypersurface of constant $u$) 
in Planck units is equal to the entropy of the non-extremal
D$p$-brane at temperature $u$. Let us compactify the 
spatial coordinates of a given horosphere $u$ on a $p$ torus
of side $L$. Then the area of a 
horosphere in Planck units is equal (up to numerical factors) to 
\be
S \sim {A \over G_N} \sim u^p L^p N^2 (N e^\f)^\g =
N^2 L^p (g_{YM}^2 N)^{(p-3)/(5-p)} u^{(9-p)/(5-p)}
\ee
This is exactly the entropy of non-extremal D$p$-branes at temperature
$T=u$\cite{KT}. In other words, the position of the horosphere 
is the UV cut-off  of the worldvolume theory. 
The fact that the relation
$u^2=U^{5-p}/g^2_{YM} N$ provides a holographic energy/distance
relation for the non-conformal cases was first derived in \cite{PP}
(generalizing the analysis of \cite{SW} for the $D3$ brane).
Here we see that the dual frame formulation provides a uniform 
treatment. 

The effective coupling of the field theory is given by 
the dimensionless combination of the Yang-Mills coupling constant
$g_{YM}^2$ with the energy scale of the problem. Therefore,
for the theory `living' on a horosphere it is equal to 
$g_{eff}^2 = g_{YM}^2 N u^{p-3}$. This can be rewritten in terms
of dual-frame variables in a quite suggestive way. 
From (\ref{Dmetric}) we see that the string length 
as measured in the dual frame is equal to 
$\a'_D \equiv l_D^2=(Ne^\f)^{2/(p-7)} \a'$.
Let us also define the `dual YM-coupling constant' to be equal 
to $(g_{YM}^D)^2 = g_s (\a'_D)^{(p-3)/2}$. Then,
\be \label{geff}
g_{eff}^2 = g_{YM}^2 N u^{p-3}=(g^D_{YM})^2 N U_D^{p-3}
\ee
where $U_D$ is the energy $U$ of the D-brane probe in 
dual-frame units, i.e. $U_D=U (l_D /l_s)$ (notice that $U_D \neq u$).

In the dual frame, strings move in a background of constant curvature
of order $1/\a'$. However, the string tension is now variable, 
$T_D \sim 1/\a_D'$. Worldsheet perturbation theory is valid when
the string tension times the curvature radius (squared) is large.
This leads to the condition $g_{YM}^2 N U^{p-3} >> 1$.
For the gauge theory description to be valid we need
the effective coupling to be small. This implies
$[g_{YM}^2 N U]^{(5-p)} << 1$. These conditions are the same
as in \cite{PP}, so their analysis carries over.
String perturbation theory is valid when the dilaton 
is small. This is always true in the large $N$ limit, except near 
$u=0$ for $p<3$ and $p=6$ or $u=\infty$ for $p=4$.

Since the solution (\ref{compac}) factorizes one can immediately 
infer the existence of a lower-dimensional gauged supergravity.
The bosonic part of the action can be easily obtained 
by reducing over the $S^{8-p}$. The fermionic part 
can also be obtained in this manner. It is easier, however, 
to use the symmetries of the configuration in order to identify the 
lower-dimensional gauged supergravity. Here we briefly summarize the
results, referring to \cite{BST} for a case-by-case discussion.
For $p=0,1$ we either find a known supergravity or 
our analysis implies the existence of certain gauged 
supergravity (in $d=2,3$ there is no classification 
of gauged supergravities). In the case the brane is related 
to one of the conformal branes (i.e. $D3, M2, M5$) 
by dimensional reduction (as for instance for $D2$)
one obtains a gauged supergravity of the type 
studied in \cite{hull}. The gauge group  
is a contraction of the gauge group of the supergravity that 
corresponds to the conformal brane. The $D6$ brane yields
the $d=8$ gauged supergravity of \cite{SS}. 
In all cases, the near-horizon D$p$-brane solution becomes one of the 
domain-wall solutions studied in \cite{lpt}. 

We finish with a brief discussion of the field theory limit of 
D-instantons\cite{OS}. This case is of particular interest 
because the corresponding $0$-dimensional gauge theory
has been conjectured to give a non-perturbative definition of type IIB
superstrings \cite{ikkt}. We find that in the field theory limit
the string frame metric of D-intantons becomes flat, but for 
finite $N$ the solution preserves only 1/2 of maximal supersymmetry.
In the $N \rightarrow \infty$ limit with $g_{YM}^2 N$ finite,
however, one obtains IIB strings in flat space with vanishing 
string coupling constant. This is in agreement with \cite{ikkt}.

\vskip0.5cm
\noindent
{\large \bf Acknowledgments}

\smallskip
\noindent
I would like thank Harm Jan Boonstra, Hirosi Ooguri and Paul Townsend for 
collaboration on the issues reported here. 
Research supported by the Netherlands Organization for Scientific Research
(NWO).


\end{document}